\documentclass[pdflatex,sn-mathphys-num]{sn-jnl}

\usepackage{graphicx}%
\usepackage{multirow}%
\usepackage{amsmath,amssymb,amsfonts}%
\usepackage{amsthm}%
\usepackage{mathrsfs}%
\usepackage[title]{appendix}%
\usepackage{xcolor}%
\usepackage{textcomp}%
\usepackage{manyfoot}%
\usepackage{booktabs}%
\usepackage{algorithm}%
\usepackage{algorithmicx}%
\usepackage{algpseudocode}%
\usepackage{listings}%

\theoremstyle{thmstyleone}%
\theoremstyle{thmstyletwo}%

\theoremstyle{thmstylethree}%

\begin{document}

\title[Article Title]{Discovery of Density Limit Disruption Induced by Core-localized Alfv{\'e}nic Ion Temperature Gradient Instabilities in a Tokamak Plasma}



\author[1]{Wei Chen*}\email{chenw@swip.ac.cn}
\author[1]{Liwen Hu}
\author[1]{Jianqiang Xu}
\author[1]{Ruirui Ma}
\author[1]{Peiwan Shi*}\email{shipw@swip.ac.cn}
\author[1]{Rui Ke}
\author[1]{Ting Long}
\author[2]{Zhiyong Qiu}
\author[1]{Haotian Chen}
\author[1]{Xiaoxue He}
\author[1]{Yonggao Li}
\author[1]{Liming Yu}
\author[1]{Wenping Guo}
\author[1]{Min Jiang}
\author[1]{Jinming Gao}
\author[1]{Xin Yu}
\author[1]{Zhengji Li}
\author[1]{Huiling Wei}
\author[1]{Deliang Yu}
\author[1]{Zhongbing Shi}

\affil[1]{\orgdiv{Center for Fusion Sciences}, \orgname{Southwestern Institute of Physics}, \orgaddress{\street{Huangjing Road}, \city{Chengdu}, \postcode{610041}, \state{Sichuan}, \country{People's Republic of China}}}



\affil[2]{\orgdiv{Institute of Plasma Physics}, \orgname{Chinese Academy of Science}, \orgaddress{\city{Hefei}, \postcode{230031}, \state{Anhui}, \country{People's Republic of China}}}



\abstract{To achieve a high energy gain, the fusion reactor plasma must reach a very high density. However, the tokamak plasmas ofen undergo disruption when the density exceeds the Greenwald density. The density limit disruption in tokamak plasmas is a mysterious barrier to magnetic confinement nuclear fusion, and hitherto, is still an unresolved issue. Over the past several years, the high density experiments with Greenwald density ratio $n_e/n_{eG}\sim1$ has been carried out using the conventional gas-puff fuelling method in HL-2A NBI and Ohmically heated plasmas. It is found for the first time that there are multiple-branch MHD instabilities in the core plasmas while $n_e/n_{eG}>0.85$. The simulation analysis suggests that the core-localized magnetohydrodynamics (MHD) activities belong to Alfv{\'e}nic ion temperature gradient (AITG) modes, and on experiment firstly, it is discovered that they trigger the minor or major disruption of bulk plasmas while the density is peaked. These new findings are of great importance to figure out and understand the origin of density limit disruptions, as well as to forecast and avoid them for future fusion rectors.}




\maketitle

\section{Introduction}\label{sec1}
Plasma disruptions are large catastrophic events in future tokamak fusion reactors. Disruptions result in a sudden confinement loss, so that heat and particles are rapidly expelled to the device wall. Disruptions can abruptly destroy the plasma facing components, and terminate the fusion reaction. Disruption issues are of central importance to future fusion rectors such as ITER\cite{Bandyopadhyay}. Meanwhile, high plasma density ($ne$) is essential for the access to high fusion gain since the fusion power density scales as $ne^2$\cite{Post}, and increasing plasma density may reduce the size of future fusion reactors and lower the cost of the devices\cite{Angioni}. However, there is a limit (known as Greenwald density limit) for tokamak high density discharges\cite{Greenwald}. The Greenwald density limit is an empirical limit for the achievable line-averaged plasma density on experiments, namely $n_{eG}=I_p/\pi a^2$, where $n_{eG}$ is the line-averaged plasma density in units of $10^{20}m^{-3}$, $I_p$ the plasma current in $MA$ and $a$ the minor radius in $m$. Generally, when the Greenwald density is reached, the bulk plasma frequently disrupts as well as the discharge halts, namely so called density limit disruptions. Density limit disruptions have been an active area of research for decades. Many previous experimental results indicate that the density limit disruption occurrence is correlated to the plasma edge cooling\cite{Rapp}, multifaceted asymmetric radiation from edge (MARFE)\cite{Pucella}, current channel shrinkage, macroscopic magnetohydrodynamics (MHD) activities (mainly tearing modes)\cite{White,Teng}, edge turbulence\cite{Long,Long2,Giacomin,Giacomin2}, and so forth. These results indicate that density limit disruptions originate from the plasma edge region\cite{Mertens,Manz,Zanca1}. Some experimental results also suggest the density limit can be exceeded by the plasma core fuelling, edge pumping, or modification of particle transport lead to peaked density profiles\cite{Lang,Bell,Esipchuk,Dingsy,Sauter,Hurst,Maingi,Liujx}. In the light of them, many theories are proposed to unravel the mechanism of density limit disruptions\cite{Gates,Singh,Zanca2,Stroth,Diamond}, however, the underlying physics is not yet fully covered and understood. In this letter, we present a new experimental evidence of density limit disruptions triggered by core-localized Alfv{\'e}nic ion temperature gradient (AITG) instabilities, and it is obviously different from existing paradigms.

\section{Results}\label{sec2}

Density limit experiments were conducted on the HL-2A tokamak\cite{Duan}. The experimental campaign involved systematic scanning of plasma density under controlled conditions, utilizing neutral beam injection (NBI) heating scenarios\cite{Chen1}. Fig.1 presents a high-density disruption event during low-power NBI heating (ShotNo.38600). The magnetic probe signals exhibit characteristic rapid changes preceding the disruption at $t=1578$ $ms$. Before the disruption, the plasma density approaches the Greenwald density with $ne/ne_G\sim1.0$. The plasma parameter evolution reveals the interplay between plasma heating, density evolution, radiation and stability limits. Fig.2 illustrates the temporal evolution of plasma instabilities in this discharge, combining measurements of the density profile and core microwave interferometer spectrogram. The observed instabilities are unstable during $t=1525-1578$ $ms$. These instabilities are rarely observed in the Mirnov signals. These instabilities, characterized by poloidal ($m$) and toroidal ($n$) mode numbers, can exhibit complex frequency dynamics including multiple spectral bands, mode coexistence, abrupt frequency jumps, and stair-like frequency progression\cite{Chen2}. These instability characteristics are determined by the following specialized diagnostics. The magnetic pickup probes exclusively measured low-frequency (LF) modes with toroidal mode numbers $n=2-4$. The core beam emission spectroscopy (BES) solely detected LF modes with poloidal mode numbers $m=4-10$. The core microwave interferometer provided an optimal response exclusively for frequencies in the $0-500$ $kHz$ range, as shown in Fig.A1. Notably, these diagnostic techniques can resolve only low mode numbers, whereas the higher mode numbers are derived from low mode number measurements and Doppler frequency shift induced by the plasma toroidal rotation. The $m/n=2/1$ instability is localized at the $q=2$ rational surface, while the high-n instabilities are localized at the $q=m/n\sim1$ surface. As the plasma density increases, these instabilities evolve from high-$n$ to low-$n$ modes while propagating from the core toward the edge.

\begin{figure}[!htbp]
\centering
\includegraphics[scale=0.7]{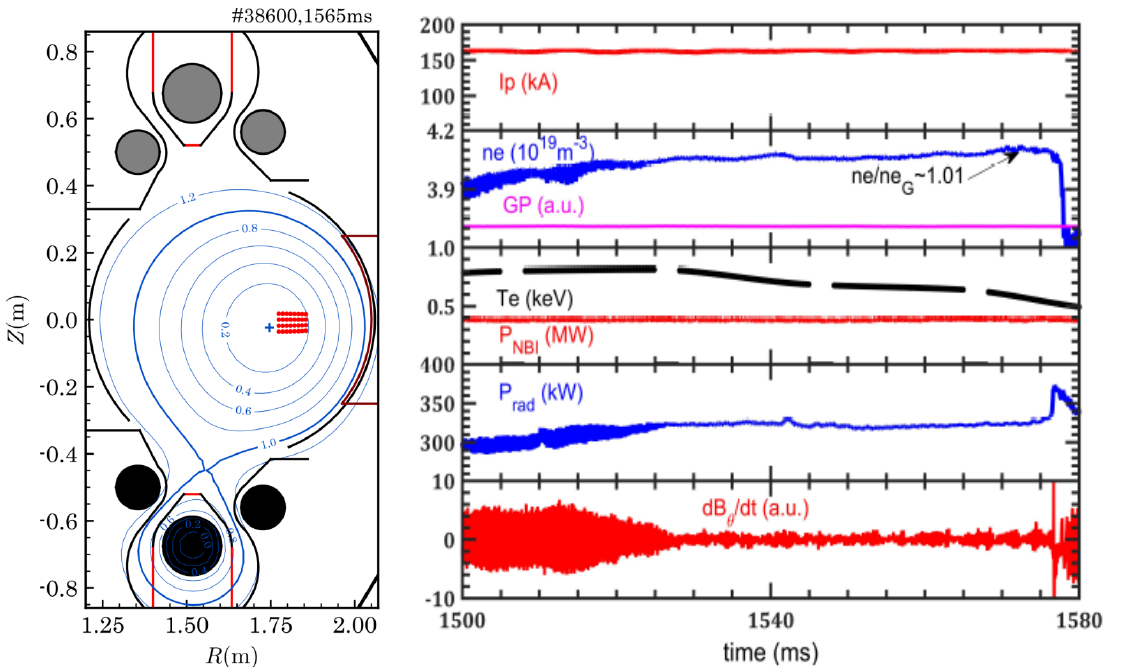}
\caption{\label{fig1} A typical high-density disruption discharge with low-power NBI heating on HL-2A (ShotNo.38600). Plasma poloidal flux surface at $t=1565$ $ms$, and the red dots show the locations of the core beam emission spectroscopy (Left col.). From top to bottom: plasma current ($I_p$), line averaged electron density ($ne$), standard gas puffing ($GP$), core electron temperature from TS ($Te$), NBI heating power ($P_{NBI}$), total radiation power ($P_{rad}$), and magnetic probe signals ($dB_{\theta}/dt$) (Right col.).}
\end{figure}

\begin{figure}[!htbp]
\centering
\includegraphics[scale=0.5]{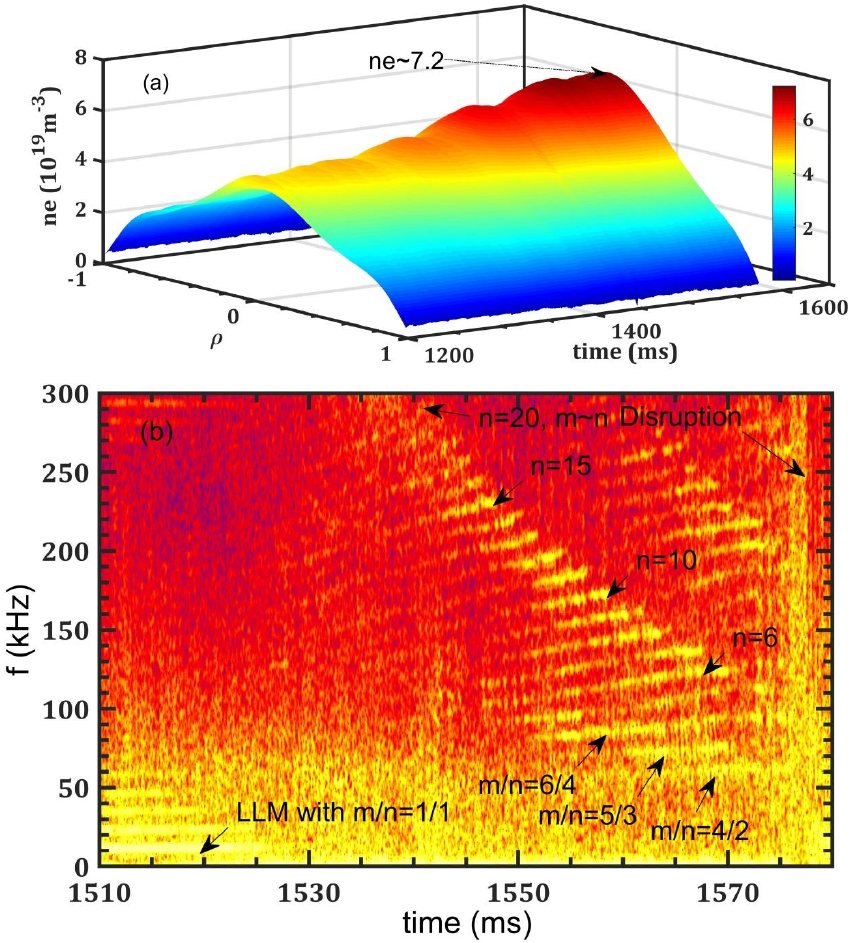}
\caption{\label{fig2}  Time evolution of the plasma density profile (a) and spectrogram of the core microwave interferometer signal (b)(ShotNo.38600). The LLM in the figure denotes the long-lived mode instability, with $m$ and $n$ being the poloidal and toroidal mode numbers of the instability, respectively. The observed instabilities during $t=1525-1578$ $ms$ exhibit the distinct frequency dynamics, such as multiple bands, mode coexistence, frequency jumps and stair-like patterns.}
\end{figure}

\begin{figure}[!htbp]
\centering
\includegraphics[scale=0.58]{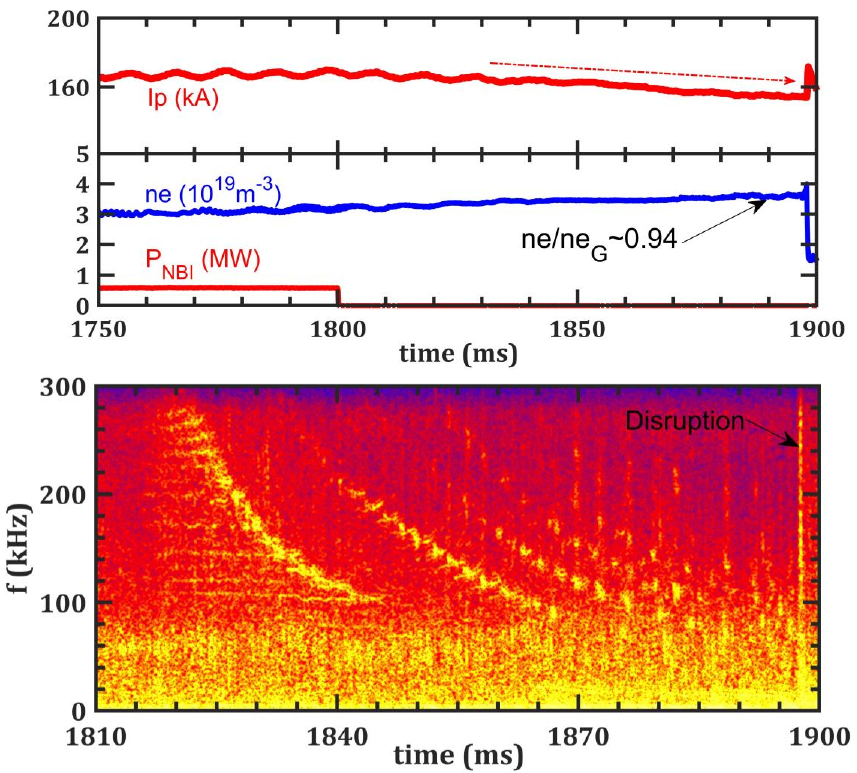}
\caption{\label{fig3} Time evolution of kinetic instabilities in high-density regime after NBI switched-off on HL-2A (ShotNo.38546). From top to bottom: plasma current ($I_p$), line averaged electron density ($ne$), NBI heating power ($P_{NBI}$), and spectrogram of the core microwave interferometer signal. The instabilities exhibit the distinct frequency dynamics, such as mode coexistence and structures resembling Christmas lights or mountain peaks.}
\end{figure}

Fig.3 presents the kinetic instability evolution in HL-2A (ShotNo.38546) following NBI switch-off in the high-density regime. These observed instabilities display complex temporal evolution patterns that vary systematically with plasma parameters and heating conditions. The observed frequency structures provide valuable signatures for real-time disruption prediction systems. These instabilities are not observed during NBI heating due to the low plasma density, but emerge after NBI heating as the plasma density increased. Since their excitation does not depend on NBI heating, these instabilities are driven unstable by energetic-ions. Remarkably, identical instabilities have also been observed in HL-2A pure Ohmic high-density plasmas. Following NBI switch-off, these instabilities exhibit distinctive frequency dynamics including mode coexistence and structures resembling Christmas lights or mountain peaks\cite{Chen2,Heidbrink}. These patterns reflect the complex nonlinear interactions between different mode structures and their coupling with background plasma profiles. These instabilities precede the disruption by $\delta t=(1892-1817)$ $ms=75$ $ms$, therefore, it can be concluded that these instabilities reveal characteristic behaviors that serve as important signatures for disruption prediction.

These measured instabilities have been identified as AITG modes by the gyrokinetic code GENE\cite{Jenko} and general fishbone-like dispersion relation (GFLDR)\cite{ChenZonca}, respectively. Linear stability analysis were performed using the GENE at different radial locations $\rho$. An analytical $s-\alpha$ equilibrium\cite{Connor} is used and it is believed that it would not affect the nature of instabilities since the HL-2A plasmas are well characterized by a circular flux surface and the geometrical effects are generally small. Fig.4(a-b) present the plasma profiles including the temperatures, density, toroidal rotation frequency and safety factor. Fig.4(c-d) are the linear eigenvalue spectrums at four radial positions. These modes are identified as AITGs, which can be inferred from their large positive real frequencies, as shown in Fig.4d. Considering the plasma toroidal rotation frequency, the simulated mode frequency shows good agreement with experimental measurements. The physical mechanism is that the magnetic shear is relatively small at the diagnostic positions hence these electromagnetic modes can be excited easily, which are further destabilized when moving outward as the gradients becomes larger. The simulations revealed that low and medium wavenumber modes become unstable across multiple radial positions, with instability growth increasing from inner to outer regions due to strengthening plasma gradients, as shown in Fig.4c. This radial dependence provides important insights into the instability drive mechanisms and their relationship with plasma parameters.

\begin{figure}[!htbp]
\centering
\includegraphics[scale=0.45]{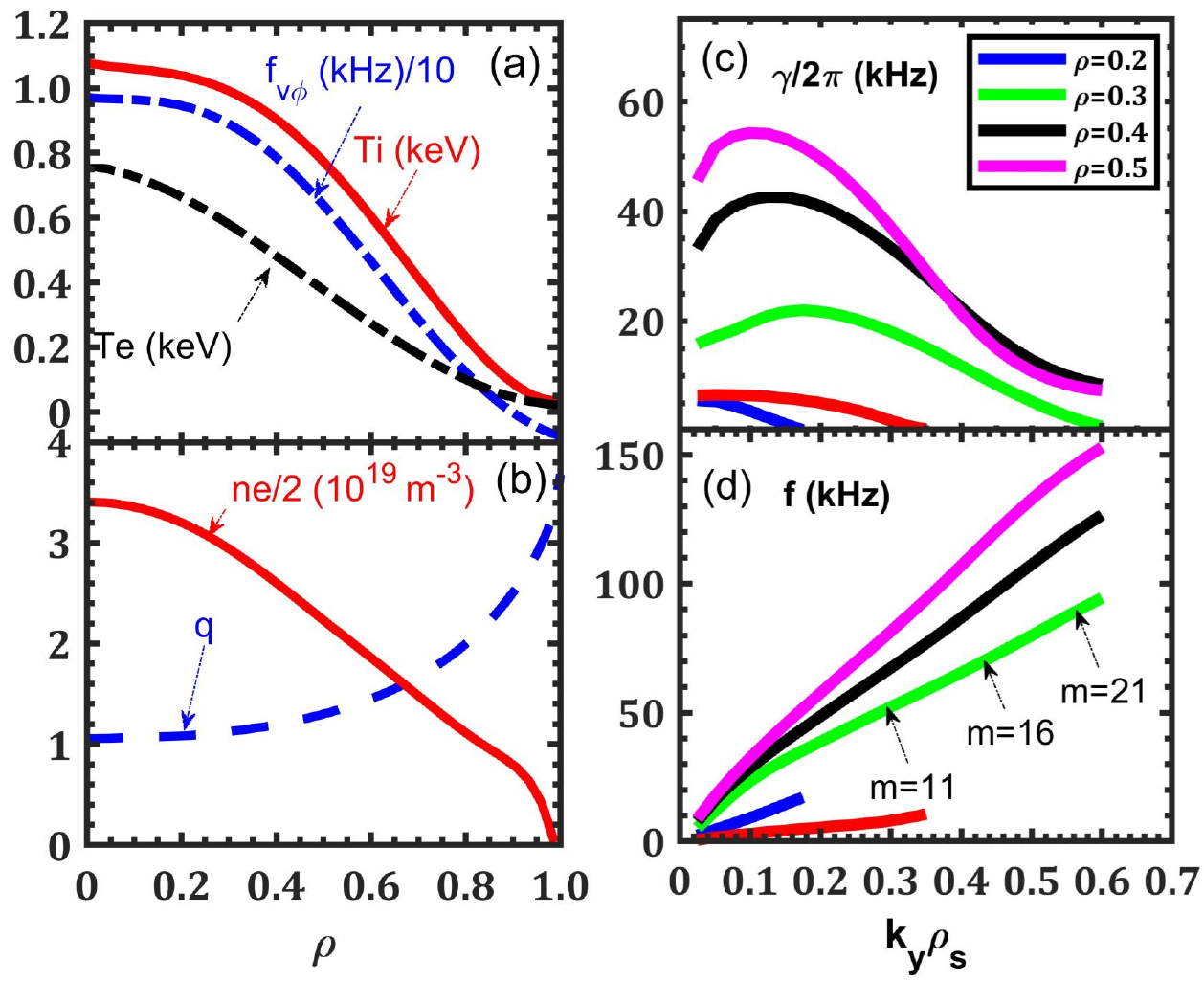}
\caption{\label{fig4} Plasma profiles (a-b), showing the electron temperature ($Te$), ion temperature($Ti$), toroidal rotation frequency ($f_{v\phi}$), electron density ($ne$), and safety factor ($q$), respectively (ShotNo.38600, t=1537 ms). Dependence of instability growth rates (c) and real frequencies (d) on wave-numbers ($k_y \rho_s$) at different radial positions ($\rho$), as given by GENE code.}
\end{figure}

 The stabilities of these modes are also analyzed using the theoretical framework of the GFLDR\cite{ChenZonca,Zonca1,Zonca2,Marr}. Based on experimental equilibrium parameters from the ShotNo.38600, our theoretical analysis indicates that for modes with the same toroidal mode number $n$, the frequency increases with the safety factor $q$, while the growth rate decreases, as clearly illustrated in Fig.5, respectively. It is important to emphasize that for a given $q$, different $n$ and their corresponding $m$ modes exhibit the following key characteristics: 1) the normalized ion temperature gradient, $\eta_i (\equiv \partial lnT_i/\partial ln n_i)$, consistently exceeds the critical threshold $\eta_{ic}$ ($\equiv (2/\sqrt{7+4\tau})(\omega_{ti}/(q\omega_{*ni}))$) with $\tau=T_e/T_i$, $\omega_{ti}$ the thermal ion transit frequency, and $\omega_{*ni}$ the thermal ion diamagnetic drift frequency due to density nonuniformity. For example, for the $n=10$ modes at $q=1.15$, $\eta_i=0.60$, while $\eta_{ic}=0.31$. 2) The MHD fluid-like contribution $\delta W_f$ in the GFLDR\cite{ChenZonca,Zonca1,Zonca2} is positive. This places the system below, yet not far below, the stability boundary of ideal MHD ballooning modes, with a magnitude on the order of $O(10^{-2})$. These findings demonstrate that the instability observed in our experiments is identified as the AITG mode\cite{Zonca99}, whose frequency is larger than thermal ion transit frequency. The AITG mode displays characteristics of dissipative-type instabilities, and it is distinct from reactive-type instabilities such as kinetic ballooning mode (KBM) or low-frequency Alfv{\'e}nic mode(LFAM)\cite{Heidbrink,Marr}. The observed increase in mode frequency with $q$ (see Fig. 5(a)) is a direct consequence of the AITG frequency scaling, $\omega \propto \omega_{*ip}$. Since $\omega_{*ip}$ increases radially with $q$, the frequency rises accordingly. The growth rate, on the other hand, is influenced by both field-line-bending stabilization and the driving term from the thermal ion pressure gradient. As the radial location moves outward and the ion pressure gradient weakens, the growth rate shows a clear decreasing trend. Moreover, for a fixed $q$, the frequency separation between modes of different n is approximately constant. This separation aligns well with five times the local plasma rotation frequency. For instance, at $q=1.1$, the frequency separation between the $n=5$ and $n=10$ modes is about $22$ $kHz$, consistent with $5f_{v\phi}=5\times4.5$ $kHz$$=22.5$ $kHz$, further validating the agreement between theoretical and experimental results. The theoretical results in Fig.5 show that by scanning the safety factor $q$, we have successfully reproduced the upward frequency shift observed experimentally. This outcome not only clarifies the physical nature of the mode but also underscores the capability of the GFLDR framework to effectively interpret and predict experimental phenomena.

\begin{figure}[!htbp]
\centering
\includegraphics[scale=0.5]{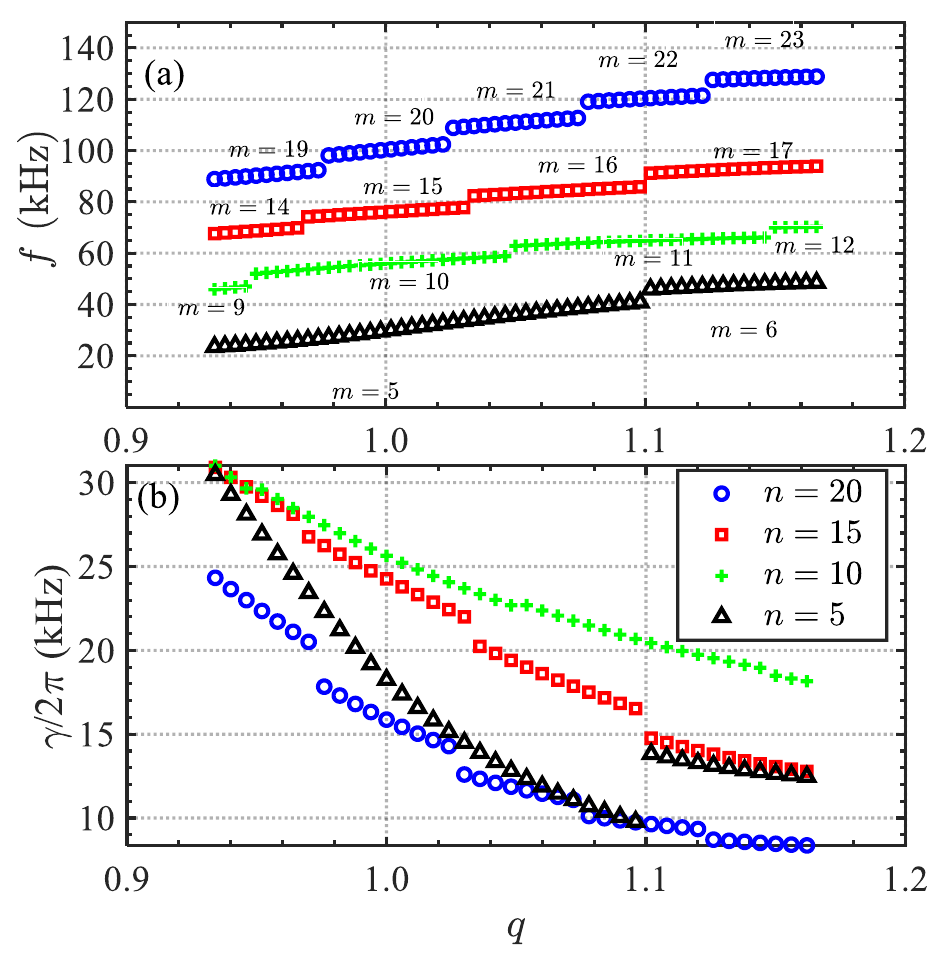}
\caption{\label{fig5} Dependence of real frequencies and growth rates of instabilities with different toroidal mode numbers on the safety factor $q$, as given by GFLDR based on experimental parameters where $\rho=0.3$.}
\end{figure}

\begin{figure}[!htbp]
\centering
\includegraphics[scale=0.55]{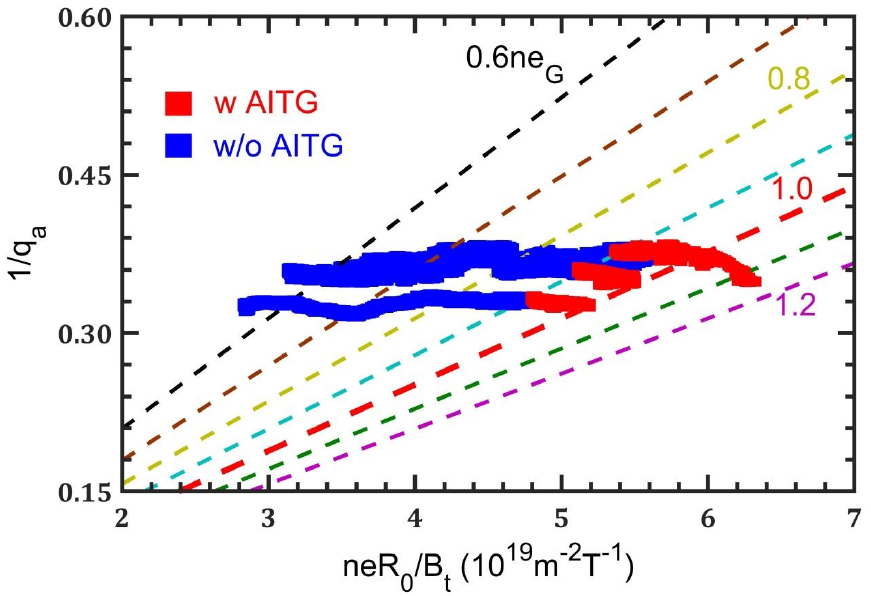}
\caption{\label{fig2} Hugill plot during the high-density NBI plasma discharges on HL-2A. $q_a=2 \pi a^2 B_t/\mu_0 R_0I_p$ is the edge safety factor, where $a$ is the minor radius, $B_t$ is the toroidal magnetic field, $R_0$ is the major radius, and $I_p$ is the plasma current.}
\end{figure}

Systematic mapping of disruption boundaries in high-density discharges reveals critical operational limits, as demonstrated by Hugill plot\cite{Hugill} analysis of HL-2A plasmas, as shown in Fig.6. It is found that there are AITG instabilities while $n_e/n_{eG}>0.85$, but they do not exist in low plasma densities. The systematic mapping of disruption points in operational space provides important insights into the stability boundaries for high-density discharges. The Hugill plot analysis during NBI plasma discharges reveals the relationship between density limit disruptions and plasma operational parameters. This analysis confirms that AITG-induced disruptions occur systematically when specific operational boundaries are crossed, providing valuable guidance for disruption avoidance strategies.

AITG instabilities are responsible for enhanced ion heat transport, resulting in the collapse of the ion temperature ($T_i$) profile. A process initiated by increasing core $T_i$ and $n_e$ excites AITG instabilities, causing a core $T_i$ drop and $T_i$-profile collapse. This process conclusively indicates that these instabilities drive strong ion heat transport(see Fig. A(2-3)). AITG instabilities exhibit a progression in which they are driven sequentially (or simultaneously) from high-$n$ to low-$n$ modes, evolving from localization at the $q=1$ to the $q=2$ surface, and from high-n small-scale to low-n large-scale structures which is a process that constitutes an inverse cascade (see Fig. A4). AITG instabilities induce outward global ballistic transport characterized by domino-like avalanches\cite{Zonca0,Yulm}, in turn, which results in the growth of the $m/n=2/1$ tearing mode (TM) and ultimately triggers a disruption (see Fig. A5). It is evident that the observed phenomenon is global, extending from the core to the edge, and electromagnetic in nature. This finding challenges all existing theoretical models of density limit disruptions. A new physics picture emerges that unifies the mechanisms leading to density limit disruptions. The system enters an unstable positive feedback cycle: initial fueling and density peaking promote edge cooling and core AITG instability, which in turn flattens the density profile and drives TM growth towards disruption, thus reinforcing the cycle. The experimental data reveal a consistent pattern wherein the plasma enters a precarious state characterized by complex MHD activity as density increases. The saturation of density evolution coincides precisely with instability onset, indicating enhanced outward radial particle transport driven by these core-localized instabilities. This correlation was observed across multiple discharges, suggesting a fundamental relationship between AITG activity and transport enhancement. In tokamak internal transport barrier (ITB) and stellarator high density regimes, the particle and heat transport caused by these kinetic instabilities had also been observed and simulated\cite{Chenw3,Duxd,Xujq,Jianx,Mulholland}. Furthermore the results suggest that they enhance plasma transport\cite{Xujq,Jianx,Mulholland}.

\section{Conclusions and Discussions}\label{sec13}
We present the recent experimental results of the density limit disruption and core-localized kinetic MHD instabilities in HL-2A NBI plasmas. It is discovered for the first time that there are multiple branch MHD instabilities in the core plasmas while $n_e/n_{eG}>0.85$. The analysis suggests that the core localized (from $\rho_{q=1}$ to $\rho_{q=2}$) MHD activities belong to AITG modes, and it is found, for the first time, in experiment that they trigger the disruption of bulk plasmas while the density is peaked. These AITG instabilities hold promise as artificial intelligence (AI) predictors for real time disruption forecasting in future tokamak fusion reactors, with predictive lead times of tens of milliseconds. The implications for next generation fusion devices are profound: traditional disruption avoidance strategies, which have focused predominantly on edge conditioning and control, may need to be augmented by core instability management approaches. Our findings establish a new physical picture that fundamentally challenges the conventional edge-centric paradigm for density limit disruptions. The identification of core localized AITG instabilities as triggers of disruptive events marks a significant advance in understanding plasma stability boundaries in toroidal confinement systems.

\section{Methods}\label{sec11}
\subsection{HL-2A Tokamak}
 The HL-2A is a medium-sized tokamak, a key device for experimental research in nuclear fusion energy in China\cite{Duan}. It represents a critical component of China's fusion research program and is operated by the Southwestern Institute of Physics (SWIP) in Chengdu, Sichuan. HL-2A has a major radius of 1.65 $m$ and a minor radius of 0.40 $m$. The toroidal magnetic field was designed as 2.8 $T$ at the magnetic axis, and the plasma current was up to 0.48 $MA$. HL-2A is equipped with a suite of advanced heating systems, including two NBl lines delivering a combined power of over 4 $MW$, with beam energies up to 40-60 $keV$. The NBI system supports tangential injection for both heating and momentum input. A rich set of diagnostics has been installed on HL-2A including charge exchange recombination spectroscopy (CXRS), Thomson scattering (TS), beam emission spectroscopy (BES), microwave interferometer, magnetic pickup probes, electron cyclotron emission (ECE), providing comprehensive measurements of plasma profiles, parameters, and MHD instability\cite{Chen1}.

\subsection{Density limit experiment setups}
A series of experiments on the HL-2A tokamak were conducted with the following typical plasma parameters: toroidal magnetic field $ B_0\sim1.2-1.4 $ T, plasma current $ I_p \sim 150 $ kA, line-averaged electron density $ n_e \sim (3.0-4.5)\times10^{19} \rm{m}^{-3} $, and edge safety factor $ q_{95}\sim 3-4$. Heating was primarily provided by co-current NBI with energies of 40-60 keV and power up to 2.5 MW.

\subsection{Mode number identifications}
Magnetic probes are fundamental diagnostic tools used to identify and characterize magnetohydrodynamic (MHD) instabilities. A key parameter they help determine is the mode number ($m$, $n$), which describes the spatial structure of an instability wave propagating in the plasma. Toroidal/poloidal mode number ($n/m$) indicates how many times the instability wave pattern repeats around the toroidal/poloidal direction of the tokamak. Beam Emission Spectroscopy (BES) is a powerful diagnostic technique used in magnetically confined fusion experiments to measure plasma density fluctuations. A key application is determining the poloidal mode number ($m=k_\theta r_d$) of plasma instabilities, which describes the number of times a wave pattern repeats poloidally.

\subsection{What is the AITG mode}
AITG modes are the electromagnetic counterparts of electrostatic ion temperature gradient (ITG) modes. The AITG mode can become unstable while the ion compression effects couple to shear Alfv{\'e}n waves (SAWs)and $\Omega_{*ip}\sim \sqrt{7/4+\tau}q$ is fulfilled, where $\Omega_{*ip}=\omega_{*ip}/\omega_{ti}$ is the ratio of ion-diamagnetic-drift and thermal-ion-transit frequencies and $\tau=T_e/T_i$ is the ratio of electron and ion temperatures\cite{Zonca1,Zonca2}. It can be understood as a branch connecting kinetic ballooning mode (KBM) (diamagnetic effects be dominant, $\Omega_{*ip}\gg\sqrt{7/4+\tau}q$) and beta-induce Alfv{\'e}n eigenmode (BAE) (ion compression effects be dominant, $\Omega_{*ip}\ll\sqrt{7/4+\tau}q$). The AITG modes can become unstable for $\eta_i$ larger than a critical value $\eta_{ic}$ which is given by $\eta_{ic}\equiv 2/(\sqrt{7+4\tau}q \Omega_{*ni})$\cite{Zonca3}, where $\Omega_{*ni}=\omega_{*ni}/\omega_{ti}$. With density peaking and a decrease in ion temperature, $\eta_{ic}$ drops,  thereby weakening the stabilizing effect and making AITG modes prone to destabilization.

\subsection{GENE code and GFLDR}
The GENE code is a powerful and highly respected tool in plasma physics. Its simulation strength lies in its ability to perform high fidelity, first principles simulations of plasma turbulence. GENE can model different regions and geometries of a fusion plasma\cite{Jenko,Gorler}. The General Fishbone-Like Dispersion Relation (GFLDR) typically has the following general structure $\Lambda(\omega)$=$\delta W_f$+$\delta W_k$\cite{ChenZonca,Zonca1,Zonca2}, where $\Lambda$ represents the generalized inertial response, while $\delta W_f$ and $W_k$ correspond to the generalized potential energy from the fluid-like plasma response and the kinetic plasmas behavior, respectively. GFLDR is a fundamental theoretical model in plasma physics used to describe the stability of certain magnetohydrodynamic (MHD) instabilities in a tokamak plasma, particularly those driven by energetic particles.

\backmatter


\section*{Declarations}


\subsection*{Acknowledgements}
We gratefully acknowledge the HL-2A team for their support in device operation, experimental execution and data collection. We thank Prof. Liu Chen, Dr. Fulvio Zonca, Prof. Zhiwei Ma and Dr. Xiang Jian for useful discussions. This work was partially supported by National Natural Science Foundation of China (Grants No.12125502 and 12535013), National Key R\&D Program of China (Grant No.2024YFE03050004), and by Innovation Program of Southwestern Institute of Physics (Grants No.202301XWCX001).

\subsection*{Author contributions}
Wei Chen, Liwen Hu, Peiwan Shi, Liming Yu, Ting Long, and Zhongbing Shi contributed to the design and execution of the HL-2A experiments. Wei Chen found and identified the phenomenon of density limit disruption induced by core-localized AITG instabilities. Wei Chen, Jianqiang Xu and Ruirui Ma performed the main data analysis. Zhiyong Qiu and Haotian Chen contributed knowledge to AITG mode and assisted with the theoretical analysis. Xiaoxue He and Deliang Yu provided the CXRS data. Rui Ke contributed BES data. Wenping Guo provided the Thomson scattering data. Zhengji Li carried out EFIT calculations. Min Jiang and Xin Yu provided the ECE data. Yonggao Li provided the plasma density data. Jinming Gao provided the plasma impurity data. Huiling Wei led the development of the NBI heating system on HL-2A. The manuscript was written by Wei Chen, with input and revisions from all co-authors.

\subsection*{Funding}
Not applicable.
\subsection*{Competing interests}
The authors declare no competing interests.
\subsection*{Ethics approval and consent to participate}
Not applicable.
\subsection*{Consent for publication}
Not applicable.
\subsection*{Data availability}
Raw data were generated from the HL-2A teams. The data supporting the findings of this work are available from the corresponding author upon request.
\subsection*{Materials availability}
Not applicable.
\subsection*{Code availability}
Source codes were developed by authors. The access can be available from the corresponding author upon request.



\bigskip
\begin{flushleft}%
Editorial Policies for:

\bigskip\noindent
Springer journals and proceedings: \url{https://www.springer.com/gp/editorial-policies}

\bigskip\noindent
Nature Portfolio journals: \url{https://www.nature.com/nature-research/editorial-policies}

\bigskip\noindent
\textit{Scientific Reports}: \url{https://www.nature.com/srep/journal-policies/editorial-policies}

\bigskip\noindent
BMC journals: \url{https://www.biomedcentral.com/getpublished/editorial-policies}
\end{flushleft}

\begin{appendices}

\vskip 20cm

\section{Extended Data}\label{secA1}






\begin{figure}[!htbp]
\centering
\includegraphics[scale=0.6]{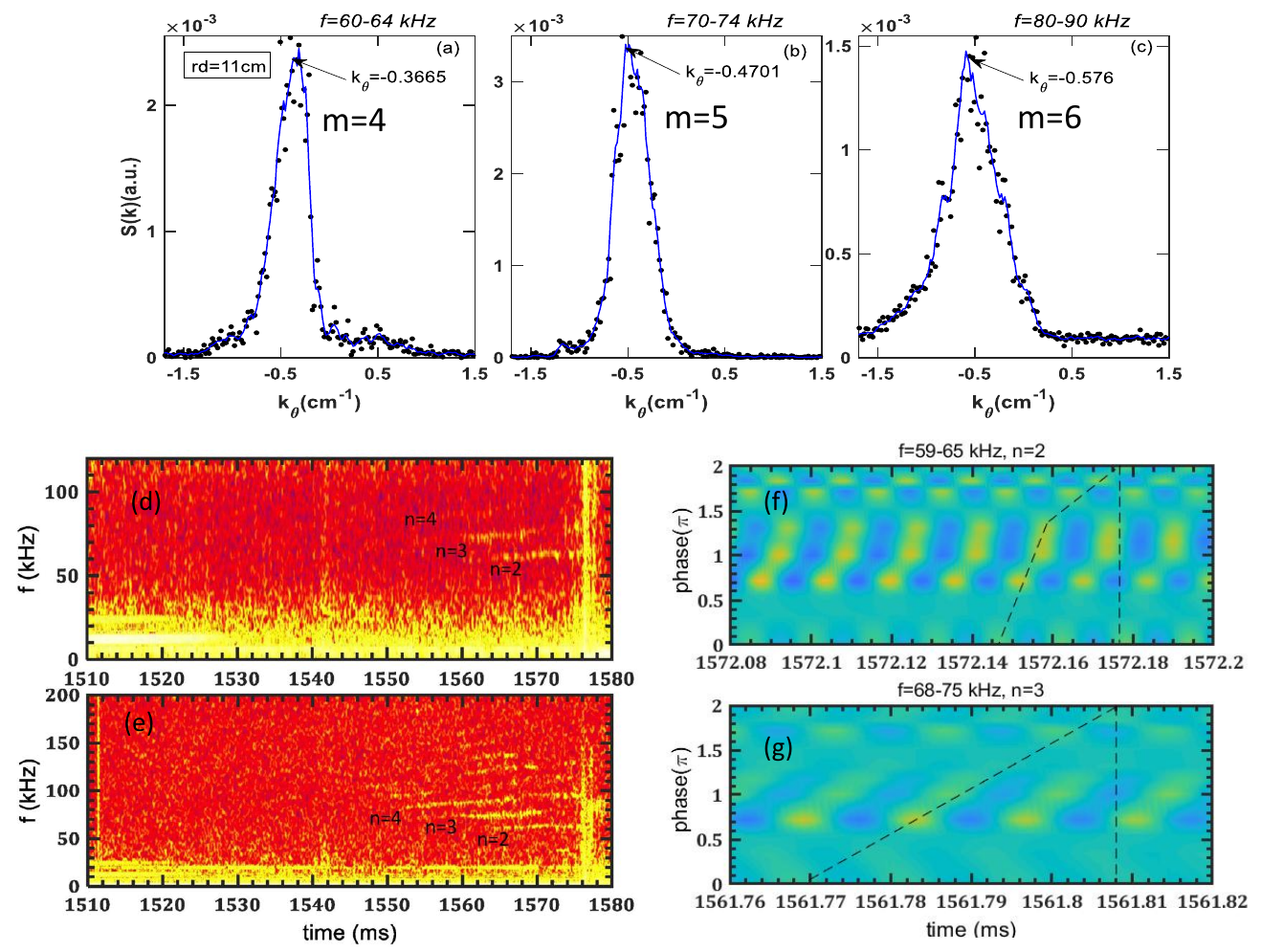}
\caption{\label{figa1} Determination of the instability mode numbers. The poloidal mode number ($m=4-6$) is identified by BES (a-c). The toroidal mode number ($n=2-3$) is determined by magnetic pickup probe signals (f-g). Spectrograms of the magnetic and BES signals are shown in (d) and (e), respectively (ShotNo.38600).}
\end{figure}

\begin{figure}[!htbp]
\centering
\includegraphics[scale=0.50]{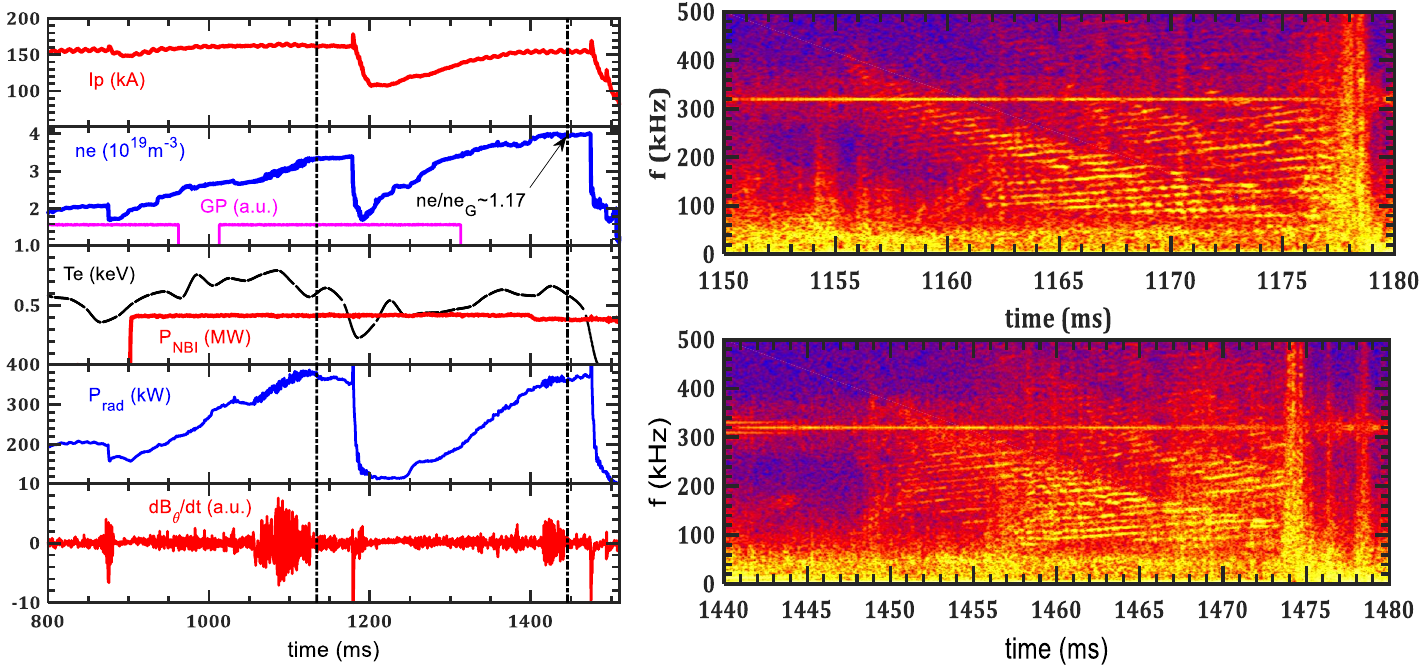}
\caption{\label{figa2} High-density discharge with multiple minor disruptions on HL-2A (ShotNo.38596). From top to bottom: plasma current ($I_p$), line averaged electron density ($ne$), standard gas puffing ($GP$), core electron temperature from TS ($Te$), NBI heating power ($P_{NBI}$), total radiation power ($P_{rad}$), and magnetic probe signals ($dB_{\theta}/dt$). Spectrogram of the core microwave interferometer signal during two different time periods. $dne/dt>0$ becomes $dne/dt\sim0$ at $t\sim1140$ $ms$ and at $t\sim1440$ $ms$ due to the excitation of AITG instabilities. It suggests that AITG instabilities induce particle transport outward in radial direction.
}
\end{figure}

\begin{figure}[!htbp]
\centering
\includegraphics[scale=0.72]{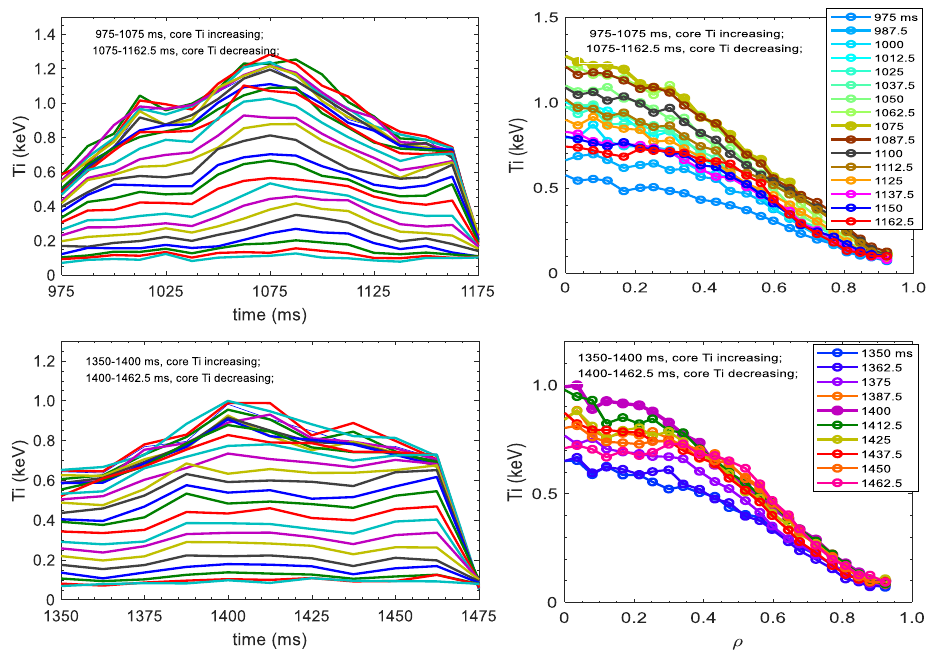}
\caption{\label{figa3} Evolution of the ion temperature and its radial profile during two different time periods (ShotNo.38546). With core Ti and ne increasing, AITG instabilities are excited, therewith the core Ti drops and Ti-profiles collapse. It indicates that these instabilities induce strong ion heat transport.
}
\end{figure}

\begin{figure}[!htbp]
\centering
\includegraphics[scale=0.8]{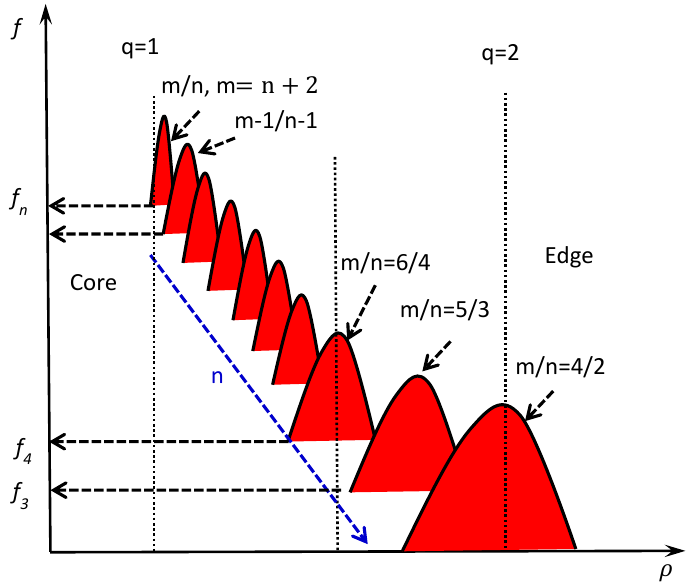}
\caption{\label{figa4} Schematic diagram of the sequential excitation of AITG instabilities according to ShotNo.38600. Ranging from high to low toroidal mode numbers and from high to low frequencies. For AITG instabilities satisfying $n=m+2$, in the limit $n\gg1$, the safety factor $q=m/n$ approaches unity ($q\sim1$), indicating high-n AITG localization near the $q\sim1$ surface.}
\end{figure}

\begin{figure}[!htbp]
\centering
\includegraphics[scale=0.5]{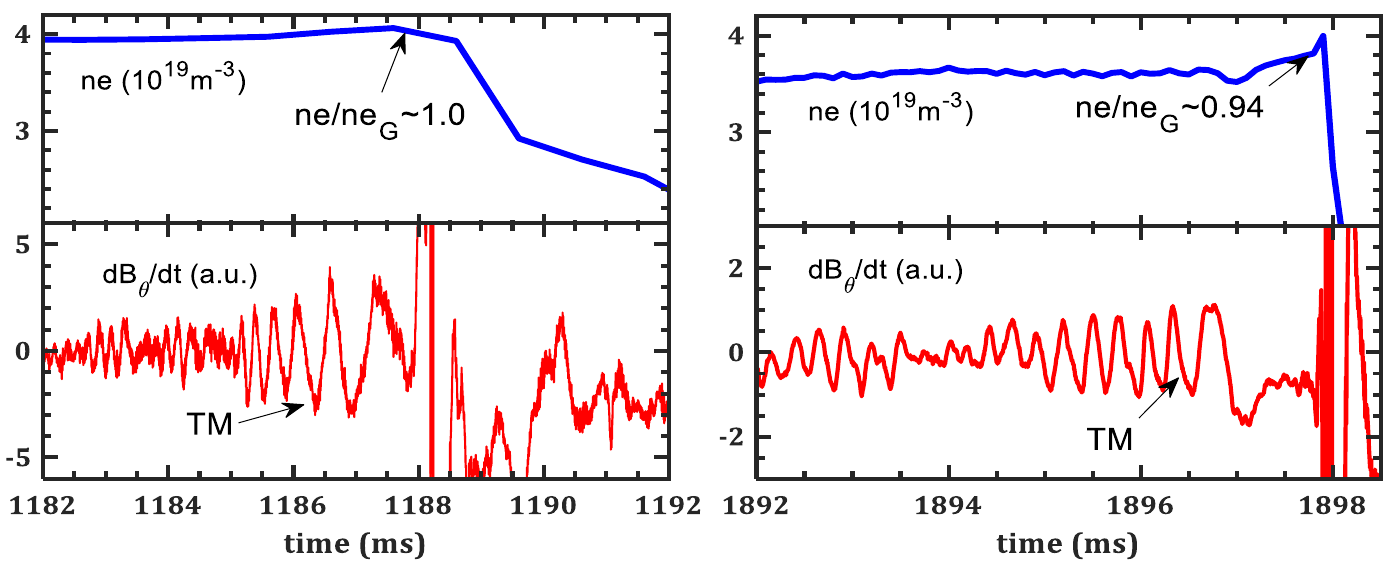}
\caption{\label{figa5} Time evolution of tearing modes (TMs) with $m/n=2/1$ before the plasma density limit disruption on HL-2A (ShotNo.39116 and ShotNo.38546). TMs are consistently detected as precursors to the disruption, manifesting a temporal lead of $\sim 5$ $ms$, which is far shorter the duration of measured AITG instabilities.}
\end{figure}

%
%

\end{appendices}



\end{document}